\documentclass{mn2e}
\usepackage{graphics}

\newcommand{\eg}{{\sl e.g.}}

\newcommand{\etal}{{\sl et al. }}

\newcommand{\Msun}{\mbox{$M_{\odot}$}}

\newcommand{\ltsimeq}{\raisebox{-0.6ex}{$\,\stackrel 
        {\raisebox{-.2ex}{$\textstyle <$}}{\sim}\,$}} 
\newcommand{\gtsimeq}{\raisebox{-0.6ex}{$\,\stackrel
        {\raisebox{-.2ex}{$\textstyle >$}}{\sim}\,$}}

\newcommand{\JHK}{$J\!H\!K~$}
\newcommand{\degs}{$^{\circ}$}

\newcommand{\ergs}{$\,$erg$\,$s$^{-1}$}
\newcommand{\Chandra}{{\it Chandra }}

\title[IR Imaging of Chandra Sources near the Galactic Centre]
{An Infrared Imaging Survey of the Faint Chandra Sources near the Galactic Centre}
\author[R.M. Bandyopadhyay et al.]{R.M. Bandyopadhyay,$^{1}$\thanks{email:rmb@astro.ox.ac.uk} J.C.A. Miller-Jones,$^{1,2}$ K.M. Blundell,$^{1}$ F.E. Bauer,$^{3}$ \and Ph. Podsiadlowski,$^{1}$ A.J. Gosling,$^{1}$ Q.D. Wang,$^{4}$ E. Pfahl,$^{5}$ and S. Rappaport$^{6}$\\ 
$^{1}$Department of Astrophysics, University of Oxford, Keble Road, Oxford, OX1 3RH, UK \\
$^{2}$Astronomical Institute ``Anton Pannekoek'', University of Amsterdam, Kruislaan 403, Amsterdam, The Netherlands, 1098 SJ\\
$^{3}$Columbia Astrophysics Laboratory, Columbia University, 550 W. 120th St., New York, NY 10027, USA\\
$^{4}$Dept. of Astronomy, University of Massachusetts, Amherst, MA 01003, USA\\
$^{5}$Dept. of Astronomy, University of Virginia, Charlottesville, VA 22903, USA\\
$^{6}$Center for Space Research, Massachusetts Institute of Technology, Cambridge, MA 02139, USA\\
}

\begin{document}

\maketitle

\begin{abstract}

\noindent
We present near-IR imaging of a sample of the faint, hard X-ray
sources discovered in the 2001 \Chandra ACIS-I survey towards the
Galactic Centre (GC) (Wang \etal 2002).  These $\sim$800 discrete
sources represent an important and previously undetected population
within the Galaxy.  From our VLT observations of 77 X-ray sources, we
identify candidate $K$-band counterparts to 75\% of the \Chandra
sources in our sample.  The near-IR magnitudes and colours of the
majority of candidate counterparts are consistent with highly reddened
stars, indicating that most of the \Chandra sources are likely to be
accreting binaries at or near the GC.
\end{abstract}


\section{The Chandra Galactic Centre Survey}

The origin and contribution of diffuse and discrete X-ray sources to
the X-ray spectrum and the total X-ray luminosity of the Galactic
Plane have been a point of debate for decades.  The unprecedented
sensitivity and angular resolution of the {\it Chandra X-ray
Observatory} has been utilized by Wang \etal (2002; hereafter W02) and
Muno \etal (2003) to investigate the nature of this Galactic Centre
(GC) X-ray emission.  The W02 ACIS-I survey of the central
0.8\degs$\times$2\degs of the GC revealed a large population of
previously undiscovered discrete weak sources with X-ray luminosities
of $10^{32}-10^{35}$\ergs.  The nature of these $\sim$800 newly
detected sources, which may contribute $\sim$10\% of the total X-ray
emission of the GC, is as yet unknown.  In contrast to the populations
of faint AGN discovered from deep X-ray imaging out of the Galactic
Plane, our calculations suggest that the extragalactic contribution to
the hard point source population over the entire W02 survey is $\leq$
10\%, consistent with the $\log(N)-\log(S)$ function derived from the
\Chandra Deep Field data (CDF; Brandt \etal 2001).  The harder
($\geq$3 keV) X-ray sources (for which the softer X-rays have been
absorbed by the interstellar medium) are likely to be at or beyond the
distance of the GC, while the softer sources are likely to be
foreground X-ray active stars or cataclysmic variables (CVs) within a
few kpc of the Sun.  The distribution of X-ray colours (Fig.~1)
suggests that $<$ 30\% of the \Chandra sources are foreground objects;
a more detailed discussion of the X-ray source characteristics will be
presented in Muno, Bauer, \& Bandyopadhyay, ({\it in prep}).

\begin{figure*}
\vspace{-0.3cm}
\scalebox{0.45}{\includegraphics{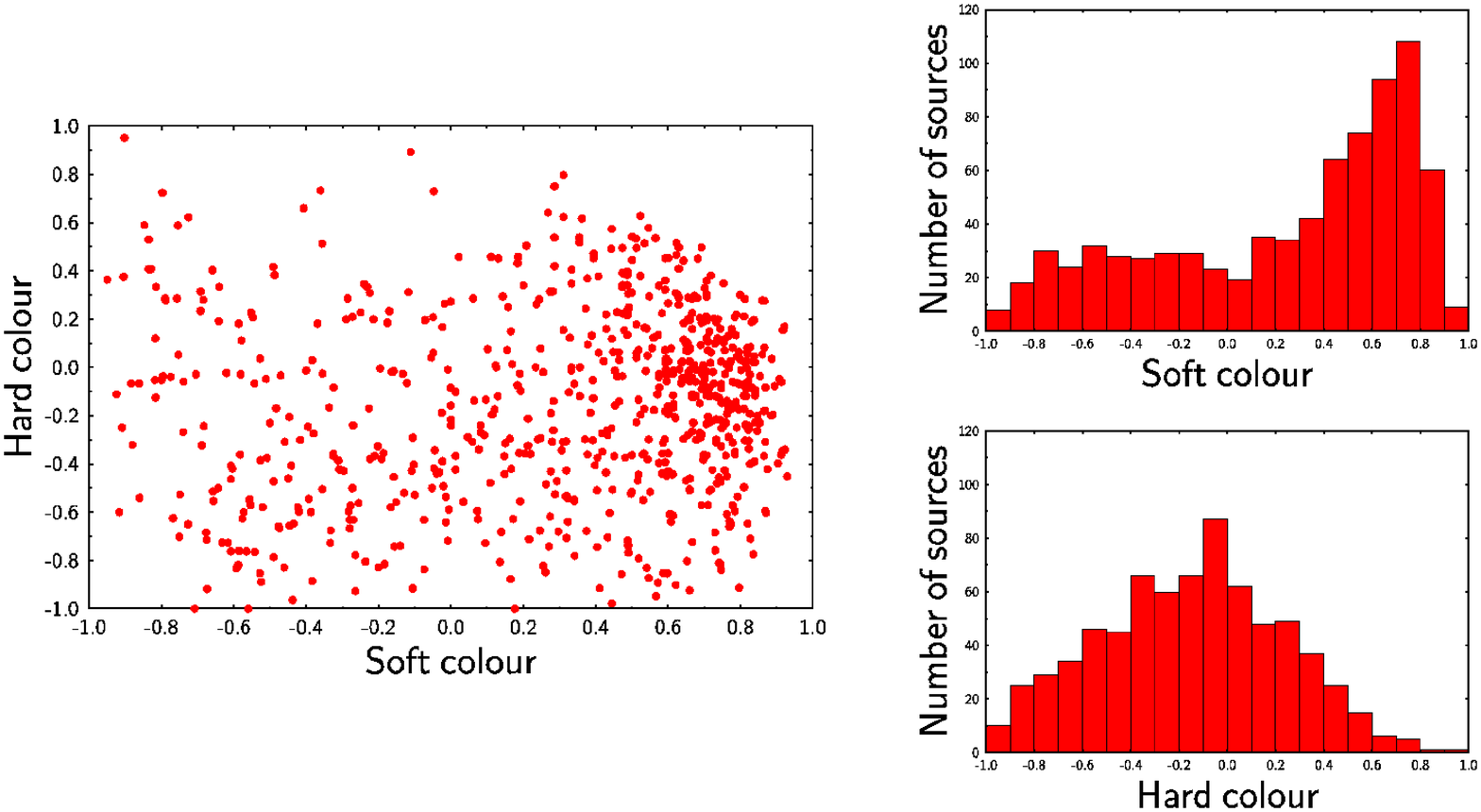}}
\vspace{-0.2cm}
  \caption{Characteristics of the X-ray source population detected in
the \Chandra mosaic.  Left: colour-colour diagram.  Right: histograms
of the number of soft and hard X-ray sources in the GC field.  The
three X-ray bands are hard (H: 5-8 keV), medium (M: 3-5 keV), and soft
(S: 1-3 keV).  Soft colour is defined as {\it (M-S/M+S)}, and hard
colour as {\it (H-M/H+M)}.  The sources which are soft and thus most
likely foreground objects are those located in the bottom left
quadrant of the colour-colour diagram. }
\label{fig:xray}
\end{figure*}

Pfahl \etal (2002; hereafter PRP02) considered in detail the likely
nature of these \Chandra sources and concluded on the basis of binary
population synthesis (BPS) models that many, if not the majority, of
these systems are wind-accreting neutron star binary systems
(hereafter WNS).  Depending on the masses of the companions, the WNSs
may belong to the ``missing'' population of wind-accreting Be/X-ray
transients in quiescence, or to the progenitors of intermediate-mass X-ray
binaries (IMXBs; 3$\leq M$/{\Msun}$\leq$7).  The existence of tens of
thousands of quiescent Be/XRBs in the Galaxy has been predicted since
the early 1980s (Rappaport \& van den Heuvel 1982; Meurs \& van den
Heuvel 1989), while it has only recently been recognized that IMXBs
may constitute a very important class of XRBs that had not been
considered before (King \& Ritter 1999; Podsiadlowski \& Rappaport
2000; Podsiadlowski \etal 2002).  In a detailed population synthesis
study of low- and intermediate-mass XRBs (L/IMXBs), Pfahl \etal (2003)
found that 80-95\,\% of all L/IMXBs should in fact descend from
intermediate-mass systems.  The W02 \Chandra survey may contain up to
10\% of the entire Galactic population of WNSs.  In addition to the
WNSs, a fraction of the \Chandra sources could be CVs (Ebisawa \etal
2005) or transient low-mass XRBs/black-hole binaries (PRP02;
Belczynski \& Taam 2004).

The first step in determining the nature of this population is to
identify counterparts to the X-ray sources.  These observations must
necessarily be done in the infrared due to the high optical extinction
in the direction of the GC.  The successful achievement of our goals
requires astrometric accuracy and high angular resolution to overcome
the confusion limit of the crowded GC.  The 2MASS survey has a
limiting magnitude of $K$=14.3, and although the astrometric positions
are accurate to 0.2\arcsec, the survey has a spatial resolution of
$\geq$2\arcsec.  As such, the 2MASS data are severely confusion
limited in the GC and moreover are of insufficient depth to detect the
majority of the expected counterparts.  We therefore constructed a
survey program using the ISAAC IR camera on one of the 8-metre
telescopes of the Very Large Telescope (VLT) at the European Southern
Observatory (ESO) in Chile, with the goal of obtaining high-resolution
\JHK-band images in order to identify a statistically significant
number of counterparts to the X-ray sources on the basis of the
\Chandra astrometry.

\begin{figure*}
\vspace{-0.15cm}
\scalebox{0.8}{\includegraphics{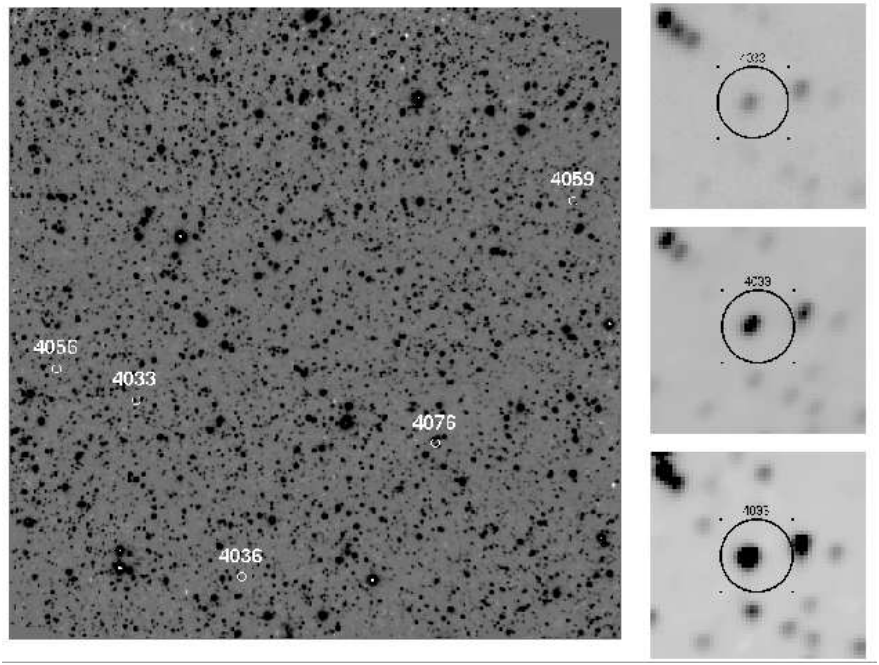}}
\vspace{-0.1cm}
  \caption{Left: an example ISAAC $K_{s}$-band field (2.5
  arcmin$^{2}$), showing the positions of 5 \Chandra X-ray sources.
  Right: Zoom-in view (8\arcsec $\times$ 8\arcsec) of the 1.3\arcsec\
  error circle of one of the \Chandra sources in this field, overlaid
  on the $J$ (top), $H$ (middle), and $K_{s}$ (bottom) ISAAC images.}
\label{fig:IRfield}
\end{figure*}

\section{IR Observations and Data Analysis} 

In constructing our VLT/ISAAC program, we preferentially selected 
hard X-ray sources from the \Chandra survey, as the soft sources are
most likely to be foreground.  For the early-type donors of the WNSs,
we expect intrinsic magnitudes of $K$=11--16, with the peak of the
magnitude distribution at $K\sim$14 (PRP02); these are therefore
readily distinguishable from the majority of late-type donors expected
in low-mass black hole or neutron star transient XRBs which generally
have $K\geq$16 in quiescence.  The average extinction towards the GC
was expected to be $A_{K}\sim$2--3 (Blum \etal 1996, hereafter B96;
Cotera \etal 2000; Dutra \etal 2003); therefore by imaging to a
magnitude limit of $K$=20 we should detect most of the WNSs.  At this
limit we could also expect to identify some counterparts to the hard
X-ray sources as AGN.  However, there are no AGN with $K\leq$17 in the
\Chandra survey of the Hubble Deep Field North of Hornschemeier \etal
(2001).  Although we cannot rule out the possibility of detecting AGN
with $K<$17, we expect these to be rare; therefore in most cases we
expect AGN to be distinguishable from the WNSs, although further
observations may be required to separate AGN from low-mass XRB
counterparts.  We further note that our IR observations indicate that
there are patches in the observed fields where the extinction even in
$K$ is substantially larger than average (as high as $A_{K}\sim 6$;
see \eg\ B96), significantly reducing our detection sensitivity in the
corresponding regions (see Section 3; a more detailed discussion of
the GC extinction is presented in Gosling, Blundell, \&
Bandyopadhyay, {\it submitted}).

In 2003, we imaged 26 fields within the \Chandra survey region with
the VLT, containing a total of 77 X-ray point sources.  We utilized
ISAAC's 1024$\times$1024 pixel Hawaii Rockwell detector, which
provides a 2.5$\times$2.5 arcmin field of view at a resolution of
0.1484\arcsec\ per pixel.  Images were obtained with
$\ltsimeq$0.6\arcsec\ seeing.  For each field, four 90-second
exposures were obtained per filter, with random offsets of
$\sim20$\arcsec\ between each exposure (using a standard ISAAC jitter
template).  Thus with a total of 6 minutes integration time per
filter, the limiting magnitudes of our survey are $J$=23 (S/N=5),
$H$=21, and $K_{s}$=20 (S/N=10).

The initial reduction of the images (including flat-fielding, removal
of bad pixels, and sky subtraction) was performed using the ESO/ISAAC
pipeline software.  We then astrometrically locked the ISAAC images to
the corresponding 2MASS images, resulting in an astrometric accuracy
of 0.2\arcsec\ for the VLT images.  IR source positions and magnitudes
were derived using the SExtractor package (version 2.3.2), which
performs well in fields which are crowded but not confusion-limited,
such as our ISAAC images (see the SExtractor manual for futher
details: http://terapix.iap.fr/IMG/pdf/sextractor.pdf).  Owing to the
crowding in the field, a Mexican hat smoothing kernel of FWHM 2.5
pixels was used over a $7\times7$ pixel grid, with the ``filtering''
option activated.  This option smooths the image before doing the
candidate source detection, resulting in a gain in the ability to
detect sources within crowded fields.  After extensive trials to
optimize detection of faint sources, the detection threshold was set
to $1.15\sigma$ per pixel above the local background (sufficiently low
to ensure we detected even the faintest sources), with a minimum area
of 3 pixels (with all 3 greater than $1.15\sigma$) required for a
detection.  The resultant source positions were then overlaid on the
IR images, each of which was inspected by eye to ensure that the
SExtractor positions corresponded to real sources in the field.  In
particular, the areas around the \Chandra error circles were carefully
inspected so that we were certain that none of the SExtractor-derived
positions were spurious, but were indeed consistent with clearly
visible IR sources.

Finally, we re-analyzed the W02 \Chandra data to estimate the
astrometric accuracy of the X-ray source positions. While the average
``out-of-the-box'' absolute astrometric accuracy for \Chandra from the
pipeline data processing is 0.6\arcsec, we note that individual
pointings can be off by up to several arcseconds (we have no way of
knowing how good the \Chandra accuracy is {\it a priori}).
Unfortunately, because of the small number of X-ray sources per
\Chandra pointing with identified optical/IR counterparts (usually
$<2$ X-ray/2MASS matches per pointing, which is insufficient to
improve the \Chandra pipeline astrometry) and the typical faintness of
optical counterparts to Galactic X-ray sources such as expected here,
we cannot establish the absolute astrometry for each tile of the
survey; in fact it is even difficult to do so for the survey as a
whole.  

\begin{figure*}
\vspace{-0.7cm}
\begin{minipage}[t]{3cm}
\scalebox{0.45}{\includegraphics{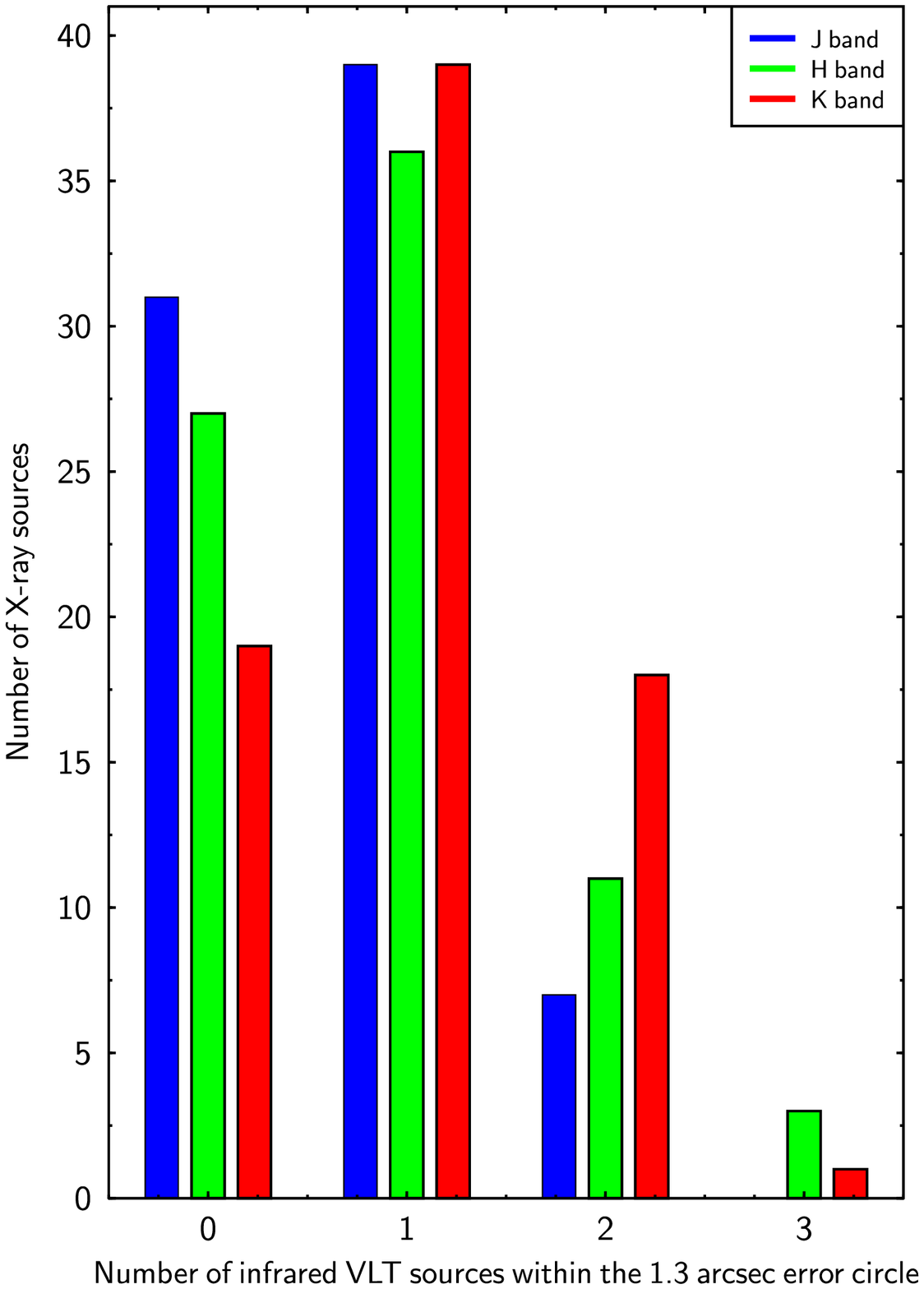}}
\end{minipage} 
\hfill
\begin{minipage}[t]{3cm}
\scalebox{0.44}{\includegraphics{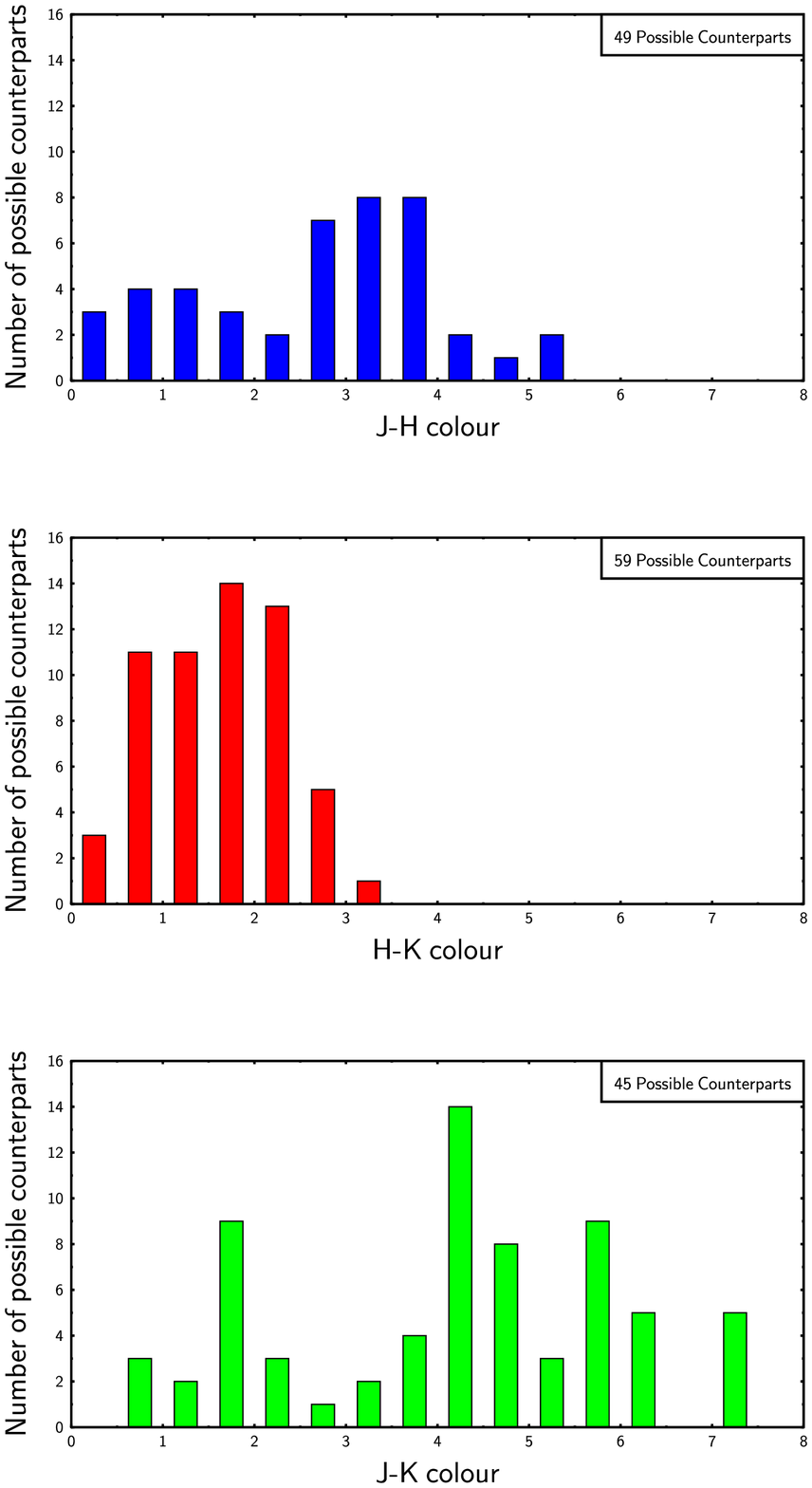}}
\end{minipage}
\hfill
\vspace{-0.2cm}
  \caption{Left: histogram of the number of candidate IR counterparts
  per \Chandra source within a 1.3\arcsec\ radius error circle.
  Right: distribution of (reddened) IR colours of potential
  counterparts.  The large spread of apparent colours is consistent
  with previously published photometry of GC stars (see \eg\ B96). }
\label{fig:counterparts}
\end{figure*}

We refined the astrometric solution for the W02 mosaic by locking it
to three significantly deeper single-pointing observations in the
\Chandra archive (each 17\arcmin$\times$17\arcmin) made within the survey
region: these fields are Sgr A* (Muno \etal (2003), Sgr B2, and the
Arches Cluster.  The Sgr A* field astrometry was originally derived
using three foreground stars, and later refined by matching 36 X-ray
sources within 5\arcmin\ of the field centre to foreground stars in the
2MASS catalog (Muno \etal 2005).  For the Sgr B2 and Arches fields,
the astrometry was established using $\gtsimeq$10 X-ray/2MASS matches
found within 5\arcmin\ of the field centre (Muno, Bauer, \&
Bandyopadhyay, 2005, {\it in prep}).  Starting with the fields
adjacent to the deep exposures and moving outward, we derived the
astrometry of the W02 mosaic by matching X-ray source positions
(obtained using the \Chandra {\it wavdetect} routine) in each pointing
to those of adjacent pointings.  However, the positional accuracy of
\Chandra X-ray sources varies strongly as a function of both off-axis
angle and source counts. Given that most of the matches between
pointings made using this method were done with X-ray sources detected
$\gtsimeq$6\arcmin\ off-axis, the statistical uncertainty of the {\it
wavdetect} positions of each source is significant.  Thus the corrected
astrometry is not substantially more accurate than the default
pipeline values, with systematic errors of up to 1\arcsec. We therefore
conservatively use an astrometric accuracy of 1\arcsec\ for the X-ray
source positions in the W02 mosaic.  For those \Chandra sources with
IR stars within a 1.3\arcsec\ radius error circle (see Section 3), the
average X-ray/IR offset is 0.77\arcsec, consistent with our assumed
\Chandra astrometric accuracy.

\section{Results: IR Characteristics}

For 75\% of the X-ray sources in our VLT fields, there are one or two
resolved $K_{s}$-band sources within a 1.3\arcsec\ error circle; only
a small number of X-ray sources have more than two potential
counterparts (Fig.~2); a complete list of all candidate IR
counterparts is presented in Table 1.  We use a radius of 1.3\arcsec\
to account for the 1\arcsec\ astrometric uncertainty of the \Chandra
data plus the uncertainty in the IR astrometry.  As the ISAAC pixel
size is $\sim0.15$\arcsec, in order to entirely cover the IR
astrometric uncertainty of 0.2\arcsec\ it was necessary to expand the
error radius by an integer number of pixels, resulting in an
additional 0.3\arcsec\ (2 pixels) being added to the 1\arcsec\ X-ray
error circle.  Over 40\% of the \Chandra sources have no potential
$J$-band counterparts, and only a few of the potential IR counterparts
have colours consistent with unreddened foreground stars (Figs.~3 and
4).  This is consistent with the expectation that the majority of the
detected X-ray sources are heavily absorbed and thus are at or beyond
the GC.

\begin{table*}
\caption{Coordinates of Candidate IR Counterparts}
\begin{center}
\begin{tabular}{llllll}\hline
\Chandra X-ray source      &  $K_{s}$-band & candidate$^{a}$  &  \Chandra X-ray source      &  $K_{s}$-band & candidate$^{a}$     \\ 
    & RA (J2000)   & Dec (J2000)       & & RA (J2000)          & Dec (J2000)      \\ \hline
CXOUJ174521.9-290519 & 17 45 21.9 & -29 05 19 &   CXOUJ174536.5-284122 & 17 45 36.6 & -28 41 21  \\
CXOUJ174518.2-290405 & 17 45 18.2 & -29 04 04 &   CXOUJ174454.2-285841 & 17 44 54.2 & -28 58 42  \\
                     & 17 45 18.2 & -29 04 06 &   CXOUJ174534.1-291327 & 17 45 34.1 & -29 13 27  \\
CXOUJ174517.2-290439 & 17 45 17.2 & -29 04 40 &   CXOUJ174457.3-290614 & 17 44 57.4 & -29 06 14  \\
                     & 17 45 17.2 & -29 04 39 &                        & 17 44 57.3 & -29 06 13  \\
CXOUJ174517.3-290625 & 17 45 17.3 & -29 06 25 &   CXOUJ174455.2-290417 & 17 44 55.2 & -29 04 17  \\
CXOUJ174503.8-290051 & 17 45 03.8 & -29 00 50 &                        & 17 44 55.2 & -29 04 16  \\
CXOUJ174608.2-290622 & 17 46 08.2 & -29 06 23 &   CXOUJ174449.9-291327 & 17 44 49.9 & -29 13 27  \\
CXOUJ174559.5-290601 & 17 45 59.5 & -29 06 02 &   CXOUJ174428.7-285651 & 17 44 28.7 & -28 56 52  \\
CXOUJ174521.9-290617 & 17 45 21.9 & -29 06 16 &                        & 17 44 28.8 & -28 56 51  \\
CXOUJ174516.8-290541 & 17 45 16.8 & -29 05 42 &   CXOUJ174459.8-291941 & 17 44 59.8 & -29 19 41  \\
CXOUJ174458.1-290509 & 17 44 58.0 & -29 05 10 &   CXOUJ174457.6-292028 & 17 44 57.7 & -29 20 28  \\
                     & 17 44 58.1 & -29 05 09 &   CXOUJ174515.8-291723 & 17 45 15.8 & -29 17 22  \\
CXOUJ174459.9-290418 & 17 44 59.9 & -29 04 18 &   CXOUJ174501.4-291933 & 17 45 01.4 & -29 19 34  \\
                     & 17 44 59.9 & -29 04 19 &   CXOUJ174450.9-291849 & 17 44 50.9 & -29 18 49  \\
CXOUJ174459.3-290050 & 17 44 59.2 & -29 00 51 &   CXOUJ174429.1-291946 & 17 44 29.0 & -29 19 46  \\
                     & 17 44 59.3 & -29 00 50 &   CXOUJ174414.3-292610 & 17 44 14.4 & -29 26 11  \\
CXOUJ174638.3-285609 & 17 46 38.3 & -28 56 09 &  CXOUJ174453.7-291952 & 17 44 53.7  &-29 19 51  \\
CXOUJ174638.7-285452 & 17 46 38.7 & -28 54 50 &                       & 17 44 53.6  &-29 19 53  \\
                     & 17 46 38.8 & -28 54 52 &  CXOUJ174429.6-291908 & 17 44 29.6  &-29 19 08  \\
CXOUJ174835.1-282336 & 17 48 35.0 & -28 23 35 &                       & 17 44 29.7  &-29 19 07  \\
CXOUJ174708.3-281410 & 17 47 08.2 & -28 14 11 &  CXOUJ174434.1-291816 & 17 44 34.2  &-29 18 16  \\
                     & 17 47 08.3 & -28 14 10 &  CXOUJ174432.1-291801 & 17 44 32.1  &-29 18 01  \\
                     & 17 47 08.3 & -28 14 09 &  CXOUJ174507.4-291557 & 17 45 07.4  &-29 15 57  \\
CXOUJ174645.2-281547 & 17 46 45.3 & -28 15 46 &  CXOUJ174354.7-290908 & 17 43 54.7  &-29 09 09  \\
                     & 17 46 45.3 & -28 15 48 &                       & 17 43 54.8  &-29 09 08  \\
CXOUJ174658.0-281414 & 17 46 58.0 & -28 14 14 &  CXOUJ174355.3-290954 & 17 43 55.3  &-29 09 54  \\
                     & 17 46 57.9 & -28 14 14 &  CXOUJ174425.9-293849 & 17 44 25.9  &-29 38 48  \\
CXOUJ174557.6-281955 & 17 45 57.6 & -28 19 56 &  CXOUJ174403.6-292742 & 17 44 03.5  &-29 27 43  \\
CXOUJ174552.4-282027 & 17 45 52.4 & -28 20 28 &                       & 17 44 03.6  &-29 27 43  \\
                     & 17 47 12.8 & -28 48 06 &                       & 17 44 03.6  &-29 27 41  \\
                     & 17 47 12.7 & -28 48 07 &  CXOUJ174407.0-292803 & 17 44 07.1  &-29 28 04  \\
CXOUJ174710.1-284931 & 17 47 10.1 & -28 49 31 &  CXOUJ174412.3-292636 & 17 44 12.2  &-29 26 36  \\
                     & 17 47 10.0 & -28 49 30 &  CXOUJ174428.5-293929 & 17 44 28.4  &-29 39 29  \\
CXOUJ174706.3-284907 & 17 47 06.2 & -28 49 07 &  CXOUJ174216.0-293756 & 17 42 16.0  &-29 37 56  \\
CXOUJ174703.1-284913 & 17 47 03.2 & -28 49 13 &                       & 17 42 15.9  &-29 37 58  \\
                     & 17 47 03.2 & -28 49 12 &  CXOUJ174216.1-293732 & 17 42 16.0  &-29 37 33  \\
CXOUJ174712.3-284828 & 17 47 12.3 & -28 48 28 &                       & 17 42 16.2  &-29 37 32  \\
CXOUJ174702.5-285258 & 17 47 02.4 & -28 52 58 &  CXOUJ174210.4-293639 & 17 42 10.3  &-29 36 40  \\
CXOUJ174536.9-284013 & 17 45 36.9 & -28 40 12 &  & & \\ \hline
\end{tabular}
\end{center}
\vspace{0.1cm}
$^{a}$Note that for some \Chandra sources there is more than one potential $K_{s}$-band counterpart.  For these X-ray sources there are multiple entries in the Table, each entry showing the coordinates for a unique $K_{s}$-band candidate counterpart to that \Chandra source.\\
\end{table*}

\begin{figure*}
\centering{
\rotatebox{-90}{
\resizebox{0.44\textwidth}{0.8\textheight}{\includegraphics*[95,-140][530,1000]{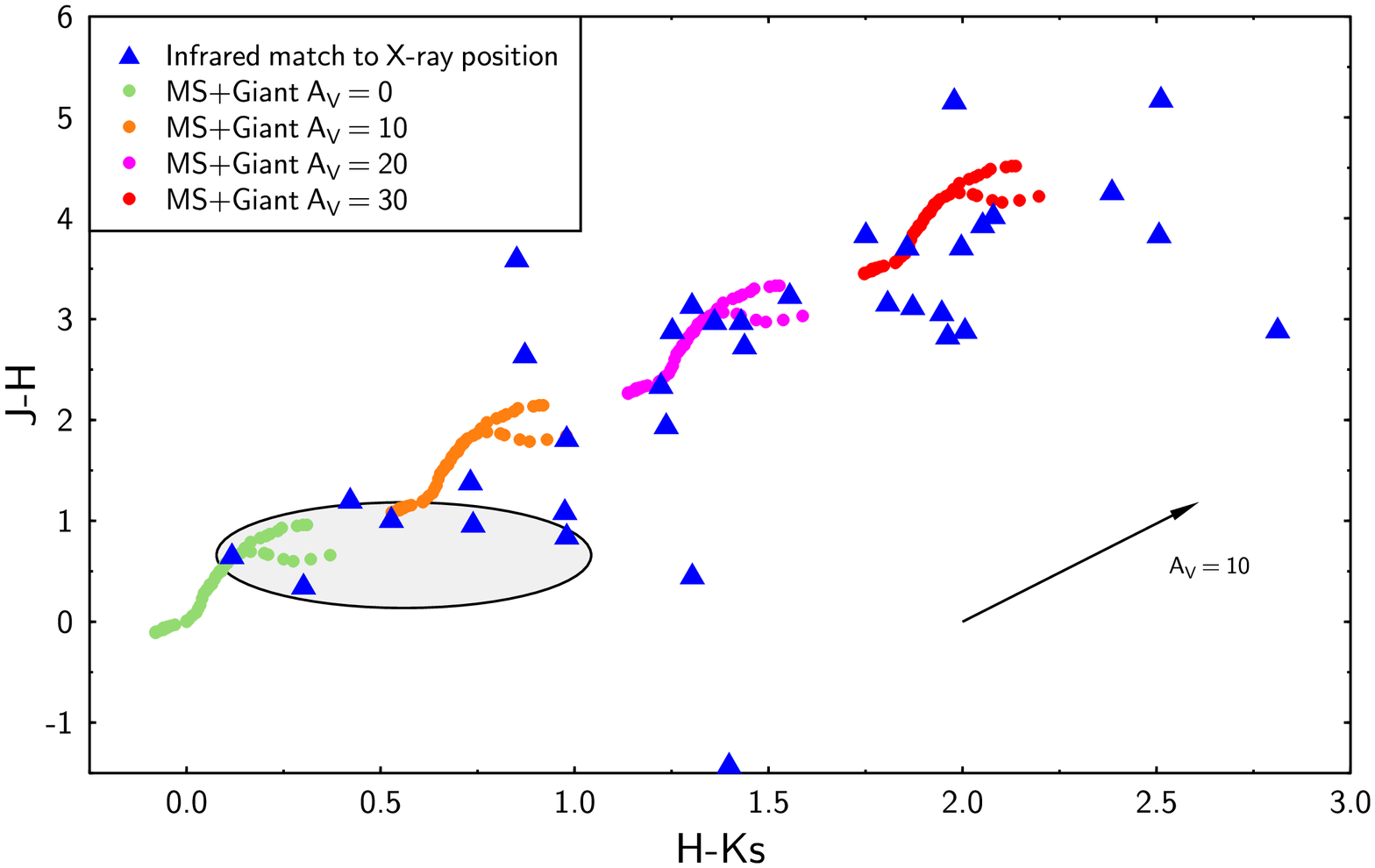}}}
\rotatebox{-90}{
\resizebox{0.44\textwidth}{0.8\textheight}{\includegraphics*[95,-140][520,1000]{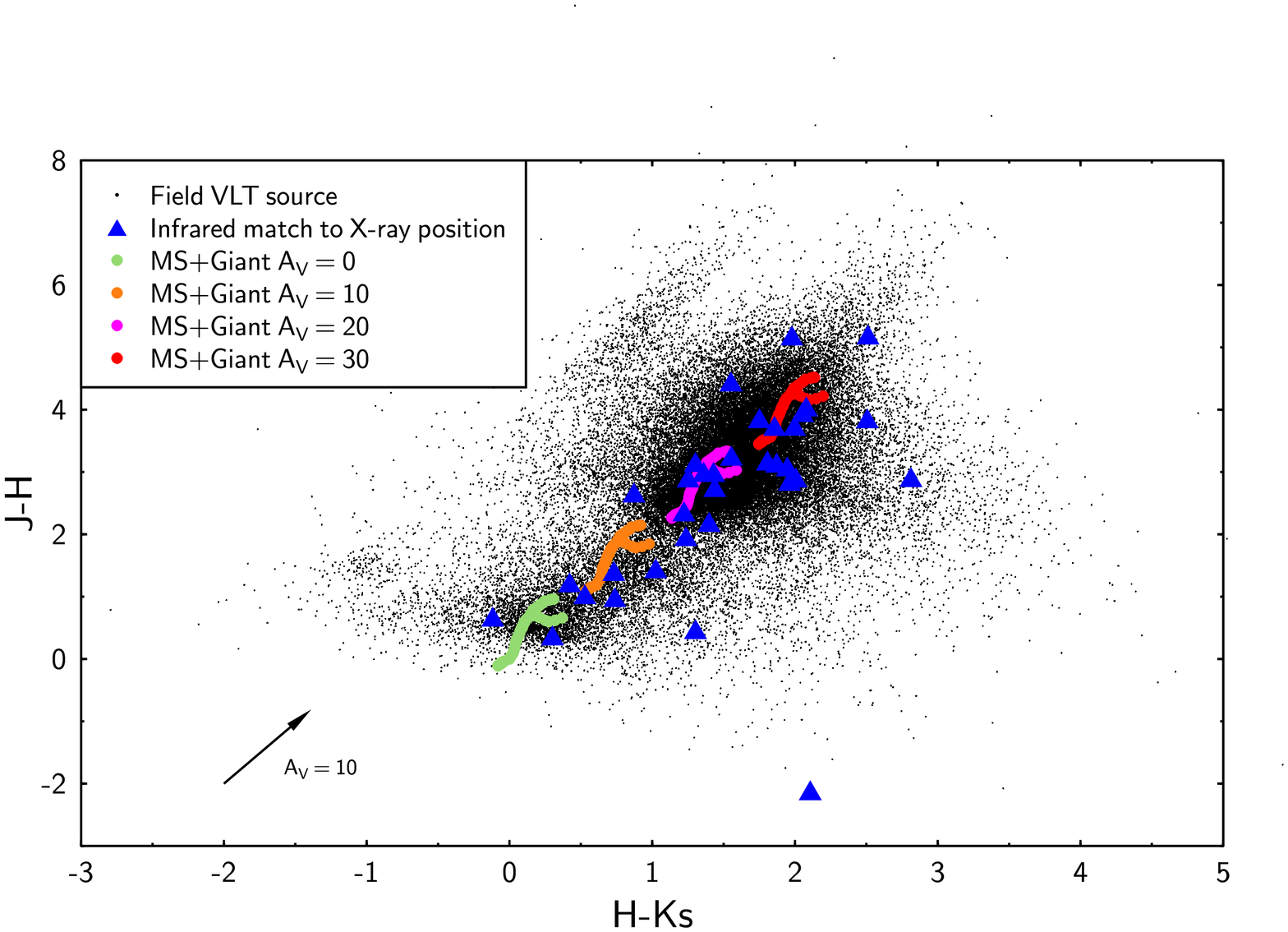}}}}
\caption{{\em Top:} Colour-colour diagram of all potential IR
  counterparts to the \Chandra sources for which we have full
  three-colour information.  The shaded oval indicates where unreddened
  AGN/QSOs would most likely be located (see text).  The theoretical
  main sequence and giant branches are indicated at visual extinctions
  of 0, 10, 20, 30 ($A_{K}/A_{V} =$ 0.11).  {\em Bottom:} Colour-colour
  diagram showing all sources in our VLT fields.  This illustrates
  that the vast majority of field stars are consistent with highly
  reddened stars; thus most of the stars (including potential X-ray
  counterparts) are at the distance of the GC (or beyond).}
\label{colour_colour}
\end{figure*}

\begin{figure*}
\rotatebox{-90}{
\scalebox{0.55}{\includegraphics*[95,-80][520,1000]{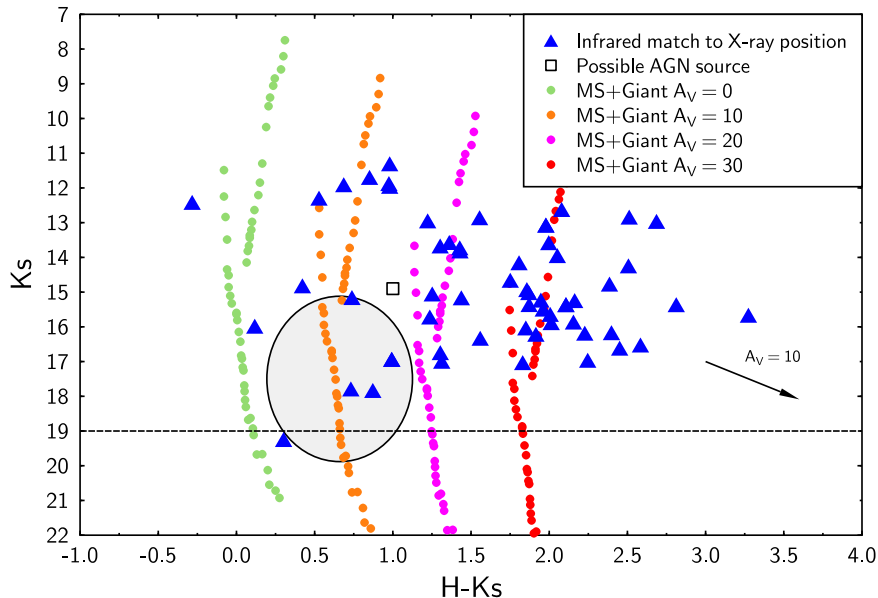}}}
  \caption{$H-K_{s}$ vs. $K_{s}$ colour-magnitude diagram for all
  candidate IR counterparts to the \Chandra sources (uncorrected for
  reddening).  Theoretical main sequence and giant branches and
  AGN/QSO locus indicated as in Fig.~3.  The small square indicates IR
  candidate counterpart for an X-ray source in the \Chandra mosaic
  whose position is consistent that of a radio source for which an AGN
  identification is claimed (Bower \etal 2001).  The dashed line shows a
  conservative estimate for the $K_{s}$ 10$\sigma$ detection limit of
  our VLT images.}
\label{fig:colour_mag}
\end{figure*}

\begin{figure*}
\vspace{-1.7cm}
\begin{minipage}[t]{3cm}
\scalebox{0.39}{\includegraphics{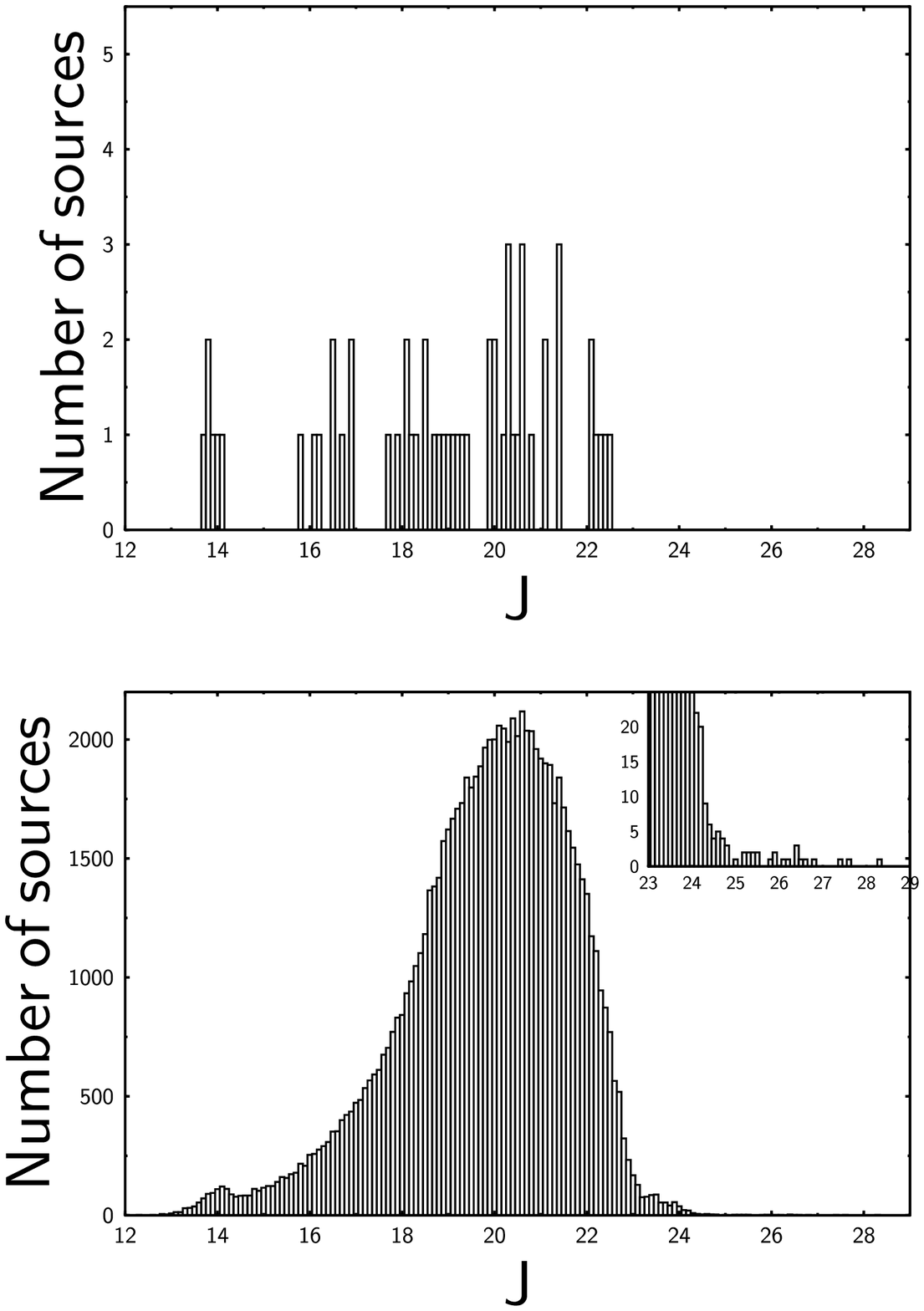}}
\end{minipage} 
\hfill
\begin{minipage}[t]{3cm}
\scalebox{0.39}{\includegraphics{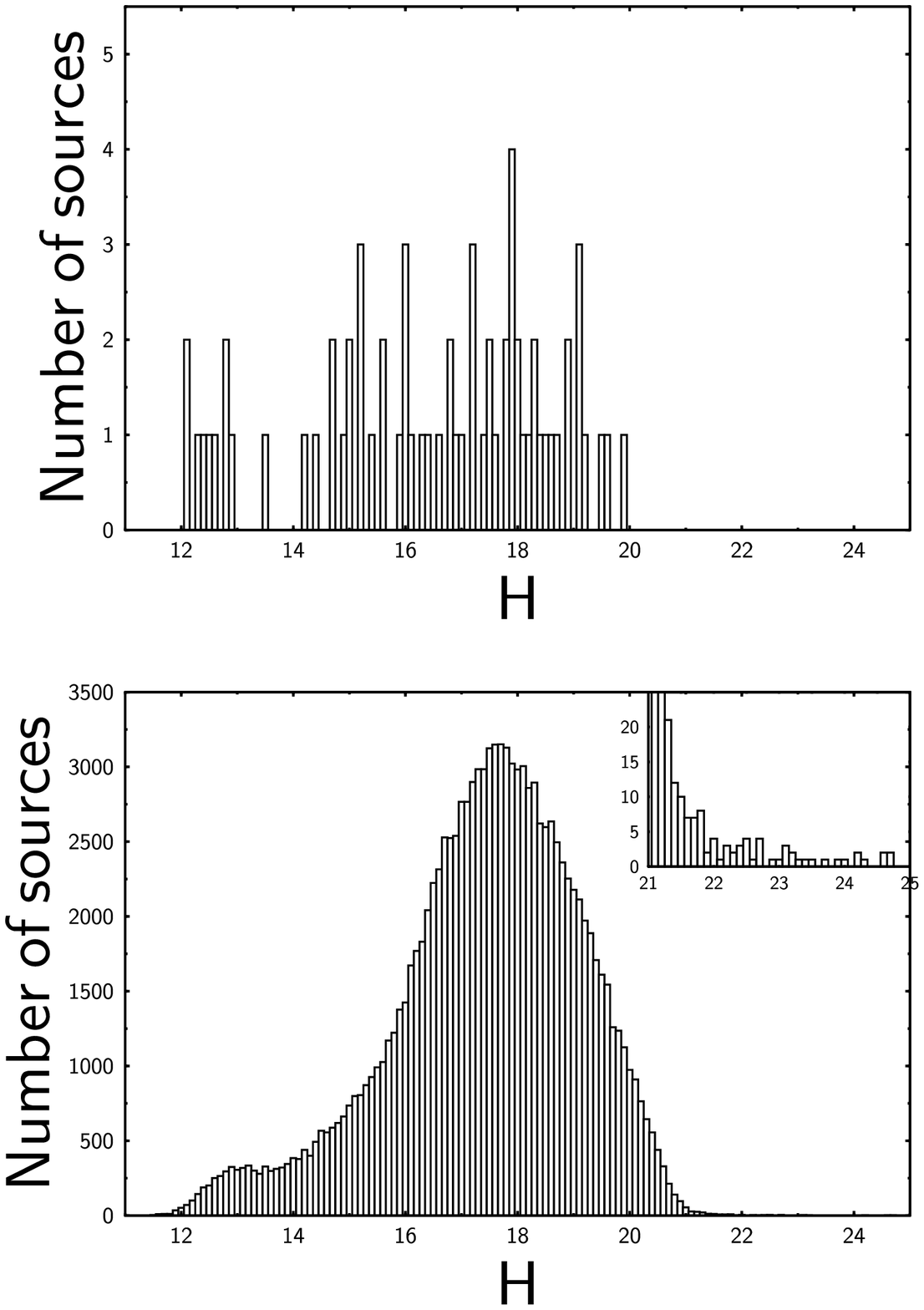}}
\end{minipage}
\hfill
\begin{minipage}[t]{3cm}
\scalebox{0.39}{\includegraphics{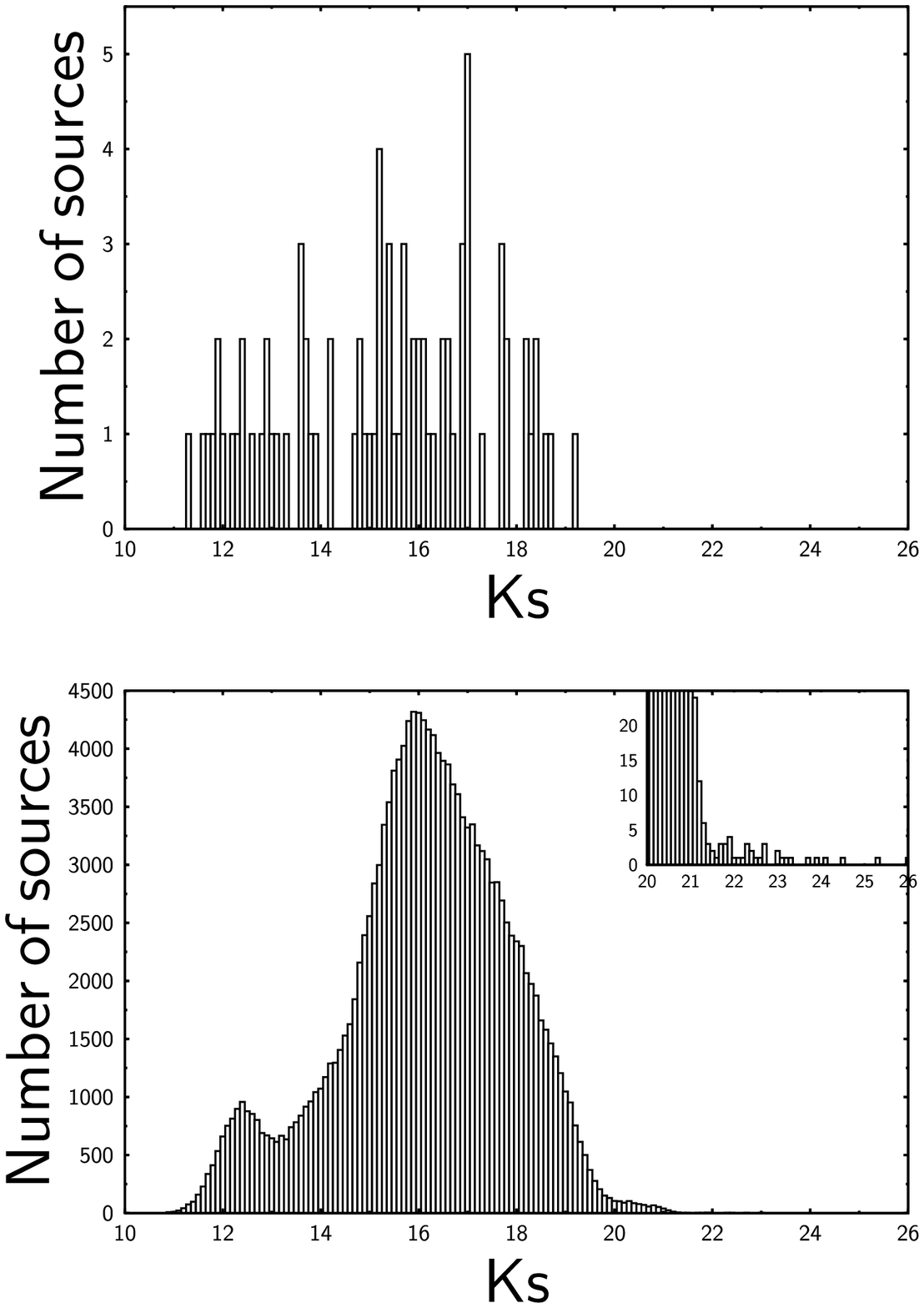}}
\end{minipage}
\hfill
\vspace{-0.3cm}
\caption{Histograms of $JHK_{s}$ (reddened) magnitudes of the
  candidate counterparts to the \Chandra\ sources (upper panels) and
  the field source population (lower panels).}
\label{fig:histograms}
\end{figure*}

For an average GC extinction of $A_{K}\sim$3.3 (B96), the peak of the
expected reddened $K_{s}$ magnitude distribution for the WNSs is
$\sim$17 (PRP02).  The peak of the observed reddened $K_{s}$
magnitudes for the potential counterparts
(Fig.~\,\ref{fig:histograms}) is $\sim$16, with an {\it (H-$K_{s}$)}
colour of $\sim$1.5--2.5 (Fig.~\,\ref{fig:colour_mag}).  There is no
obvious difference between the distribution of the \JHK magnitudes of
our candidate counterparts and those of the field population (Fig.~5);
a more detailed comparison will be presented in a subsequent paper
(Gosling \etal, {\it in prep}).

In order to determine the significance of the IR matches, a Monte
Carlo simulation was carried out to see what fraction of
randomly-located 1.3\arcsec-radius circles would contain an X-ray
source simply by chance.  The \Chandra X-ray source positions were 
shifted by a random amount between 5\arcsec\ and 20\arcsec, and the
number of shifted error circles containing an IR counterpart was
recorded. This procedure was carried out $10^4$ times, and the results
were binned according to the fraction of shifted positions whose error
circles did not contain an IR source.  The distribution obtained was
then fitted with a normal distribution.  The mean fraction of
randomised positions with non-detections inside a 1.3\arcsec\ error
circle was found to be $0.253\pm0.048$, $0.405\pm0.053$, and
$0.507\pm0.053$ for $K_{s}$, $H$, and $J$ bands respectively
(1$\sigma$ errors).  For comparison with the values obtained for the
real data as reported below, these fractions correspond to approximately
25\% (19 sources), 40\% (31 sources), and 50\% (39 sources) without
counterparts in the simulated data, for $K_{s}$, $H$, and $J$ 
respectively.  Owing to the decreasing extinction with
wavelength, the fraction of 1.3\arcsec\ circles without IR
counterparts decreased on moving from $J$ through to $K_{s}$.

Out of a total of 77 X-ray source positions in our VLT fields, the
percentage without IR counterparts in $K_{\rm s}$, $H$, and $J$
respectively were 25\% (19 sources), 35\% (27 sources), and 40\% (31
sources; see Fig.~\,\ref{fig:counterparts}), corresponding to 0.1-,
1.0- and 2.0-$\sigma$ below the expected values for a random
distribution of X-ray error circles.  These values are consistent with
what is expected as a result of the stellar density in our fields.
The average separation of stars in the $K_{s}$-band is 1.94\arcsec, thus
with a 1.3\arcsec\ error circle we would not expect a statistically
significantly larger number of counterparts than for a random
distribution.  However, in the $H$- and $J$-band, where the average
stellar separation is 2.29\arcsec\ and 2.72\arcsec\ respectively, we do
expect to see a larger number of X-ray/IR matches than would be
expected by chance, and indeed, that is what is observed.
At the shorter wavelengths, we are detecting more sources within the
X-ray error circles than would be expected by chance, and therefore we
are likely to be detecting some of the true IR counterparts to the
X-ray sources.

The detection of a larger number of potential counterparts in $J$ and
$H$ (8 and 4 more sources, respectively, than in the simulated data)
than would be expected from a random distribution has two potential
interpretations.  First, we may be detecting more matches of X-ray
sources to foreground (rather than GC) stars, as foreground stars are
subject to much less extinction.  The second possibility is that there
are a larger number of potential counterparts which are intrinsically
brighter and bluer than the average field population, a result which
would be consistent with the hypothesis that many of these faint
\Chandra sources may be WNS systems.  In any case, this simulation
shows that some fraction of the potential counterparts within the
error circles are likely to be chance coincidences, especially at $K$,
so follow-up spectroscopy must be used to identify which IR sources
are the true counterparts to the X-ray sources.

There are no $K_{s}$-band counterparts for $\sim$25\% of the \Chandra
sources.  This is larger than the expected fraction of background AGN
from the CDF estimate, though other groups have predicted larger
fractions (up to 50\%).  However, the extinction in the GC is
extremely variable, even at $K$.  From visual inspection of our VLT
data it is clear that some areas exhibit larger than average
extinction ($A_{K}\geq5$) in the form of dust patches and lanes
(Gosling, Blundell, \& Bandyopadhyay, {\it submitted}).  Seven of our
X-ray sources are in regions of unusually heavy IR (notably $K$-band)
extinction; thus we estimate that the true fraction of \Chandra
sources without counterparts at the limiting $K_{s}$ magnitude of our
survey is $\sim$16\%.  As has been noted by previous authors (B96;
Cotera \etal 2000), the highly variable and structured nature of the
extinction in the GC makes derivation of reddening-corrected
photometry for the field population non-trivial.  Thus a detailed
assessment of the likely spectral types for the candidate counterparts
will be presented by Gosling \etal ({\it in prep}).

\section{Discussion}

The colours and magnitudes of the IR candidate counterpart population
are generally consistent with main-sequence and giant stars at
distances of the GC.  Only a few sources have low extinctions
consistent with foreground stars.  Of the detected candidate
counterparts, only a few have colours and magnitudes consistent with
AGN/quasars; however these are indistinguishable from foreground stars
on the basis of magnitudes and colours alone.  From visual inspection
of our VLT images, none of the candidate counterparts appears
intrinsically extended (though this of course does not rule out QSO or
faint AGN).  In Figs.~4 and 5 we have indicated the likely colours
and magnitude range for AGN/QSOs (as derived from Spinoglio \etal
1995, Ivazic \etal 2001, Sharp \etal 2002, and Pentericci \etal 2003).
Note that for substantial visual extinction ($A_{V}\geq$10) most
extragalactic counterparts to the \Chandra sources will be below the
detection limits (Fig.~5) of our survey.  The limiting magnitude of
this survey also places limits on the types of stars detectable at the
distance of the GC.  Along the line of sight out to the distance of
the GC, in our images we expect to detect all evolved (supergiant and
giant) stars, but we will only detect main sequence stars of type G or
earlier; K, M dwarfs in the GC will be below our detection limit.

On the basis of this astrometry and photometry, we were awarded VLT
time in the summer of 2005 to obtain ISAAC $K$-band spectra of the 31
best candidate counterparts ($K_{s}$ magnitudes $\sim$12--17); the
goal of these observations is to identify the X-ray source
counterparts via detection of accretion signatures.  The primary
accretion signature in the $K$-band which distinguishes a true X-ray
counterpart from a field star is strong Brackett $\gamma$ emission;
this technique of identifying XRB counterparts has been verified with
observations of several well-studied GC XRBs (see \eg\ Bandyopadhyay
\etal 1999).  As these \Chandra sources are weaker in X-rays than the
previously known population of Galactic XRBs, and thus have lower
inferred accretion rates, their emission signatures will likely be
somewhat weaker than in the more luminous XRB population.  However,
the Brackett $\gamma$ accretion signature is clearly detected in the
IR spectra of CVs, which are only weak X-ray emitters with a similar
X-ray luminosity range to the \Chandra sources (see Dhillon \etal 1997
for IR emission signatures in CVs).  Of course, the magnitude limits
of our VLT imaging preclude detecting IR counterparts to any CVs which
may be present in our fields located beyond a few kpc from the Sun.
The comparison between CVs and the counterparts of the \Chandra
sources at GC distances is made purely to illustrate that even for
X-ray binaries accreting at lower rates than in ``canonical'' systems,
we expect the Brackett $\gamma$ signature to be detected and thus the
spectroscopic identification to be definitive even for these
low-luminosity X-ray sources.  The spectra we obtain of our brighter
targets will allow us not only to identify the counterpart via its
emission signature but also to spectrally classify the mass donors if
absorption features are detected, a crucial step in determining the
nature of this new accreting binary population.

Identifying IR counterparts to these newly discovered X-ray sources
provides a unique opportunity to make a census of the various
populations of accreting binaries in the GC and may ultimately allow a
determination of each system's physical properties.  As this \Chandra
survey may contain up to 10\% of the entire population of accreting
binary systems in the Milky Way (PRP02), our results will have important
implications for our understanding of XRBs in the Galaxy, including
their formation, evolutionary history, and physical characteristics.
In particular, if many of the detected sources do indeed turn out to be
the missing progenitors of L/IMXBs, it would put strong constraints on
models of this important population of X-ray sources.

\section{Acknowledgments}
The authors would like to thank Niel Brandt for helpful discussions
about \Chandra data analysis and astrometric issues, and Michael Muno
for information about the astrometry of the \Chandra image of Sgr A*.
K. M. B. thanks the Royal Society for a University Research
Fellowship.  J. C. A. M.-J. and A. J. G. thank the UK Particle Physics
and Astronomy Research Council for Studentships.  We also thank Mark
Morris for his helpful and rapid referee report.  This paper is based
on observations made with the ESO VLT at Paranal under program ID
71.D-0377(A).


\begin{thebibliography}{}

\bibitem{} Bandyopadhyay, R.M., Shahbaz, T., Charles, P.A., \& Naylor, T., 1999, MNRAS, 306, 417

\bibitem{} Belczynski, K. \& Taam, R.E., 2004, ApJ, 616, 1159

\bibitem{} Blum, R.D., Sellgren, K., \& DePoy, D.L., 1996, ApJ, 470, 864 (B96)

\bibitem{} Bower, G.C., Backer D.C., \& Sramek, R.A., 2001, ApJ, 558, 127

\bibitem{} Brandt, W.N., \etal, 2001, AJ, 122, 2810

\bibitem{} Cotera, A.S., Simpson, J.P., Erickson, E.F. \& Colgan, S.W.J., 2000, ApJSS, 129, 123

\bibitem{} Dhillon, V.S., \etal, 1997, MNRAS, 285, 95

\bibitem{} Dutra, C.M., Santiago, B.X., Bica, E.L.D., \& Barbuy B., 2003, MNRAS, 338, 253

\bibitem{} Ebisawa, K., \etal, 2005, ApJ, {\it in press} (astro-ph/0507185)

\bibitem{} Gosling, A.J., Blundell, K.M., \& Bandyopadhyay, R.M., {\it submitted}.

\bibitem{} Hornschemeier, A.E., \etal, 2001, ApJ, 554, 742

\bibitem{} Ivazic, Z., \etal, 2001, Proc. IAU Symposium 184, p. 137

\bibitem{} King, A.R. \& Ritter, H., 1999, MNRAS, 309, 253

\bibitem{} Meurs, E.J.A. \& van den Heuvel, E.P.J., 1989, A\&A, 226, 88

\bibitem{} Muno, M.P., \etal, 2003, ApJ, 589, 225

\bibitem{} Muno, M.P., Lu, J.R., Baganoff, F.K., Brandt, W.N., Garmire, G.P., Ghez, A.M., Hornstein, S.D., Morris, M.R., 2005, ApJ, {\it in press} (astro-ph/0503572)

\bibitem{} Pentericci, L., \etal, 2003, A\&A, 410, 75

\bibitem{} Pfahl, E., Rappaport, S., \& Podsiadlowski, Ph., 2002, ApJL, 571, L37 (PRP02)

\bibitem{} Pfahl, E., Rappaport, S., \& Podsiadlowski, Ph., 2003, ApJ, 597, 1036

\bibitem{} Podsiadlowski, Ph. \& Rappaport, S., 2000, ApJ, 529, 946

\bibitem{} Podsiadlowski, Ph., Rappaport, S., \& Pfahl, E., 2002, ApJ, 565, 1107

\bibitem{} Rappaport, S. \& van den Heuvel, E.P.J., 1982, Proc. IAU Symposium 98, p. 327

\bibitem{} Sharp, R.G., Sabbey, C.N., Vivas, A.K., Oemler Jr., A., McMahon, R.G., Hodgkin, S.T., \& Coppi, P.S., 2002, MNRAS, 337, 1162

\bibitem{} Spinoglio, L., Malkan, M.A., Rush, B., Carrasco, L., \& Recillas-Cruz, E., 1995, ApJ, 453, 616

\bibitem{} Wang, Q.D., Gotthelf, E.V., \& Lang, C.C., 2002, Nature, 415, 148 (W02)

\end{thebibliography}
\end{document}